\documentstyle[12pt]{article}
\def\a{\alpha}
\def\b{\beta}

\def\d{\delta} 			\def\D{\Delta}
\def\e{\epsilon}

\def\g{\gamma} 			\def\G{\Gamma}
\def\h{\eta}

\def\l{\lambda} 		\def\L{\Lambda}
\def\m{\mu}
\def\n{\nu}
 			
\def\p{\pi} 			
\def\q{\theta} 			
\def\r{\rho}
\def\s{\sigma} 			\def\S{\Sigma}

\def\ms2{m_{B^*}^{2}}
\def\fr{\frac}
\def\ba{\begin{array}}
\def\ea{\end{array}}
\def\bz{\begin{equation}}
\def\ez{\end{equation}}
\def\by{\begin{eqnarray}}
\def\ey{\end{eqnarray}}

\def\nn{\nonumber}
\newtoks\slashfraction
	\slashfraction={.13}
\def\slash#1{\setbox0\hbox{$\, #1$}
	\setbox0\hbox to \the\slashfraction\wd0{\hss \box0}/\box0}

\begin{document}
\begin{flushright}
FERMILAB-Conf-95/281-T
\end{flushright}
\begin{center}
\vskip 1cm
 \Large\bf Potential for Discoveries in Charm Meson Physics
\footnote{Talk delivered at the Workshop on the Tau/Charm Factory,
Argonne June
21-23 1995.}
\end{center}
\vskip 1cm
\begin{center}
\large Gustavo Burdman
\end{center}
\vskip 0.5cm
\begin{center}
\small\em Fermi National Accelerator Laboratory, P. O. Box 500,
Batavia, IL
60510, USA.
\end{center}
\vskip 1.5cm
\begin{quote}
\small{\bf Abstract.}  The possibility of using charm meson physics
to test the
Standard Model (SM) is reviewed. In the case of $D^0-\bar{D}^0$
mixing, the SM
contributions are carefully considered and the existence of a window
for the observation of new physics is discussed. Some examples of
extensions
of the SM giving large mixing signals are presented. Finally, some
distinctive
aspects of CP violation and rare decays in charm mesons are
discussed.
\end{quote}
\vskip 1cm
The $D$ meson has been largely overlooked as a testing ground for the
SM. The
reason for this might be traced back to the fact that the most
important
effects in Flavor Changing Neutral Currents (FCNC) and in flavor
mixing are
brought about, in the SM, by the top quark. This has important
effects in loops
that couple to external down quarks ($B$ mixing, radiative and rare
$B$ and $K$
decays).
In the SM, top quark loops do not couple to external up quarks and
thus the SM
loop  effects in charm physics are expected to be very small. This is
the case,
for instance, for $D^0-\bar{D}^0$ mixing and for rare decays: the SM
predicts
very small rates.  However this can be viewed as a window of
opportunity for
observing effects coming from new physics at higher energy scales. Of
these,
 the  most interesting ones are those that are relatively small or
even
negligible
  in $B$ and $K$ physics but become observable contributions when
looking at  $D$ physics. Here we will discuss where these
opportunities are
likely to be and some of the new physics scenarios giving interesting
signals.
We will first discuss $D^0-\bar{D}^0$ mixing in the SM, focusing on
our current
understanding of the long distance contributions. This is a crucial
point  in
 order to establish whether or not there is a window to observe new
physics in
the measurement of this effect. Then we will go on to show some
examples of
extensions of the SM that could fill this experimental
window. Among these are multi-Higgs doublet models, fourth-generation
effects,
supersymmetry, and tree-level FCNC effects induced by dynamical
symmetry
 breaking scenarios.
We will then move to briefly review the prospects of CP violation
effects in
$D$ physics, both direct and associated with mixing, in the SM and
beyond.
Finally, we will take a look at radiative and rare decays and point
out the
relevance of some modes as tests
of the SM.
\vskip 1cm
\begin{center}
\large\bf
$D^0-\bar{D}^0$ MIXING IN THE STANDARD MODEL
\end{center}
The current experimental knowledge of $D^0-\bar{D}^0$ mixing comes
from the
upper bound on the wrong-sign to right-sign ratio
\bz
r_D\equiv \fr{\G(D^0\to \ell^- X)}{\G(D^0\to \ell^+ X)}\simeq
\fr{1}{2}\left[
\left(\fr{\D m_D}{\G}\right)^2  + \left(\fr{\D
\G_D}{\G}\right)^2\right],
\label{rd_def}
\ez
with the approximation in (\ref{rd_def}) valid for $\D m_D/\G,\;\D
\G_D/\G\ll
1$. From the latest E691 data \cite{e691}
we know $r_D<3.7\times 10^{-3}$. If $\D\G_D/\G$ is neglected, this
translates
into an upper limit for the mass difference giving
\bz
\D m_D^{\rm exp.}<1.3\times 10^{-13}{\rm GeV} \label{exp_lim}
\ez

\begin{center}
\large\bf Short Distance
\end{center}
In the SM the short distance $\D C=2$ transition occurs via the box
diagrams.
The effective interactions at the $m_c$ scale are described by the
 hamiltonian \cite{datta}:
\bz
{\cal H}_{\rm eff.}^{\D
C=2}=\fr{G_F}{\sqrt{2}}\fr{\a\,|V_{cs}^{*}V_{us}|^2}{8\p\
sin^2\q_W}\;\fr{(m_{s}^{2}-m_{d}^{2})^2}{m_{W}^{2}m_{c}^{2}}\left(\cal
 O +\cal
O'\right) \label{eff_hal}
\ez
with ${\cal O}\equiv \bar{u}\g_\m(1-\g_5)c\;\bar{u}\g^\m(1-\g_5)c$
and
${\cal O}'\equiv \bar{u}(1+\g_5)c\;\bar{u}(1+\g_5)c$. The presence of
the
additional operator $\cal O'$ is due to the non negligible external
momentum.
The matrix elements of the operators can be parametrized by
\bz
\langle D^0 |{\cal O}|\bar{D}^0\rangle = \fr{8}{3}m_D f_{D}^{2} B_D
\quad
;\quad
\langle D^0 |{\cal O}'|\bar{D}^0\rangle =
-\fr{5}{3}\left(\fr{m_D}{m_c}\right)^2 m_D f_{D}^{2} B'_D
\label{mat_par}
\ez
In the vacuum insertion approximation  one has  $B_D=B'_D=1$ and the
short
distance contribution to the mass difference is
\bz
\D m_{D}^{\rm SD}\simeq 2.5\times 10^{-17}\;{\rm
GeV}\left(\fr{m_s}{0.3 {\rm
GeV}}\right)^4\; \left(\fr{f_D}{f_\p}\right)^2 \label{dmd_sd}
\ez
Thus for typical values of $f_D$ and $m_s$ the short distance
contributes
to $r_D$ with a value not above $10^{-8}$, perhaps as small as
$10^{-10}$.

\begin{center}
\large\bf Long Distance
\end{center}
The contributions of the short distance box diagrams are not the only
ones. The fact that light quarks with rather large CKM couplings to
the charm
quark can propagate between the $D^0$ and $\bar{D}^0$,hints the
possibility
of relatively important long distance contributions to mixing.
The propagating degrees of freedom are hadrons rather than quarks.
The
situation is very different in $K$ and $B$ mixing, where there is
always a
 very important effect of a heavy quark inside the box diagram loop,
with large
CKM couplings: the charm quark in $K^0-\bar{K}^0$ and the top quark
in
$B^0-\bar{B}^0$. In the latter, the effect of the top quark
completely
dominates and long distance contributions are expected to be
negligibly small
due to the small CKM couplings of the $B$ meson to light hadrons. In
the case
of $K$ mixing, the coupling to light-hadron intermediate states is
still large
as a consequence of which sizeable long distance contributions -of
the same
order of magnitude of the short distance ones- are expected
\cite{dgh84}.
 The long distance contributions to $D^0-\bar{D^0}$ mixing are
inherently
nonperturbative and cannot be calculated from first principles. It is
however
of paramount importance to estimate their size in order to understand
to origin
of a possible observation of the effect in future experiments.

A first observation is that the mass difference is a $SU(3)$ breaking
effect.
On the other hand, $\D  m_D$ is doubly Cabibbo suppressed whereas
$\G$ is not.
Therefore, a naive estimate of the effect would be given by
\bz
\fr{\D m_D}{\G}\sim \l^2\times (SU(3) {\rm ~breaking}) \simeq
(10^{-3}-10^{-2})
 \label{nai_est}
\ez
where $\l\simeq\sin\q_c$. Specific calculations tend to give smaller
results.
There are basically two ways to attempt estimating the long distance
effects:
a dispersive approach and Heavy Quark Effective Theory (HQET).

\begin{center}
\large Dispersive Approach
\end{center}

\begin{figure}
\vspace{4.5cm}
\includegraphics{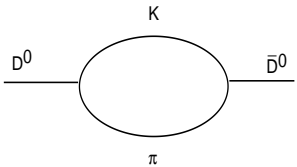}
\vspace{-1cm}
\caption[]{ Long distance contribution from two charged pseudo-scalar
intermediate states.}
\label{pp_cont}
\end{figure}

An estimate of the long  distance contributions can be obtained
 by assuming they come from the propagation of hadronic states to
which both
$D^0$ and $\bar{D}^0$ can decay. There will be one, two, three, etc.
particle
intermediate states. Each of these groups can be further separated
into sets
whose contributions vanish separately in the $SU(3)$ limit. One of
these sets
is
 formed by  the two-charged-pseudoscalar intermediate states
$\p^+\p^-$,
$K^+K^-$, $K^-\p^+$ and $K^+\p^-$.
 Thus computing
their contribution to the mass difference, as shown schematically
in Fig.~\ref{pp_cont}, gives a  concrete realization of the estimate
in
(\ref{nai_est}) for an $SU(3)$ set for which data is available.
This was first done in \cite{dght86}.

\noindent Although these ``self energy" diagrams will depend on the
interaction
chosen for the vertices,  they have a universal imaginary part which
typically
comes from a
logarithm. This leads to the expression \cite{dght86}
\bz
\fr{\D m_{D}^{\em l.d.}}{\G}\simeq
\fr{1}{2\p}\;\ln\fr{m_{D}^{2}}{\m^2}\left[B(\p^+\p^-) +
B(K^+K^-)-2\;\sqrt{B(K^-\p^+)\,B(K^+\p^-)}\right] \label{pp_ld}
\ez
 where all the branching ratios correspond to $D^0$ decays. The scale
$\m$ is a
typical hadronic scale, ${\cal O}( 1{\rm~GeV})$.
Inserting the current available experimental values \cite{pdg94}
 in (\ref{pp_ld}) we obtain
\bz
\fr{\D m_D}{\G}\simeq 8.4\times
10^{-4}\times\left(1.46-\sqrt{b}\right)
\label{ld_esp}
\ez
where we have defined $b$ by
\bz
 \fr{B(K^+\p^-)}{B(K^-\p^+)}=b\;\tan^4\q_c \label{def_b}
\ez
The experimental measurement of the Doubly Cabibbo Suppressed Decays
(DCSD) gives $b=2.8\pm 1.3$.
Then, for any  value of $b$ in a $1\s$ interval, it is clear that the
cancellation in this $SU(3)$ set is even better than expected in
(\ref{nai_est}), giving $|\D m_{D}^{\rm l.d.}/\G|\simeq {\cal
O}(10^{-4})$.

Several assumptions are implied in (\ref{pp_ld}). First, the minus
sign is
obtained assuming there is no
strong relative phase between the two $K\p$ amplitudes. This is only
true in
the $SU(3)$ limit \cite{wolf95}.
 However, the effect of this phase is expected to be small
\cite{brow_pak}.
 More importantly, equation
(\ref{pp_ld}) is valid for massless particles in the loop and it
assumes a
constant coupling at the vertices which allows one
to relate the product of the two $\D C=1$ interactions in
Fig.~\ref{pp_cont} to
the actual decay amplitudes.
The approximation regarding the internal masses is rather safe in the
two-pseudoscalar (PP) case. However masses should be kept in the
calculation
when
sets including vector mesons (PV and VV) are considered.  On the
other hand,
large momenta in the loop should not contribute given that the
coupling is
expected to develop a momentum suppression at a typical hadronic
scale. This
suggests the existence of a physical cutoff for the integrals
involved. It is
instructive to see
how this cutoff and the effect of the internal masses come about. The
contribution
of Fig.~\ref{pp_cont} to the mass difference obeys a dispersion
relation of the
form
\bz
\S(p^2)=\frac{1}{\p}\;\int_{s_0}^{\infty} \frac{Im[\S(s)]\; ds}{(s-
p^2-i\e)} \label{dis_rel}
\ez
where $s_0\equiv (m_1+m_2)^2$, and
$m_1$ and $m_2$ are the masses in the loop. Taking into account a
subtraction
forcing the condition $\S(0)=0$ and keeping the masses, the
implementation of a
cutoff $\L$ in the dispersive integral (\ref{dis_rel}) gives
\cite{bghp2}
\by
\fr{\D m_{D}^{\rm l.d.}}{\G}&\simeq&\frac{m_D}{4\p}\left\{
\frac{B(\p^+\p^-)}{{\bf
p}_{\p\p}}\;I(m_\p,m_\p,\L)+
\frac{B(K^+K^-)}{{\bf p}_{KK}}\;I(m_K,m_K,\L) \right. \nn \\
& &\left. \quad\quad -2\fr{\sqrt{B(K^-\p^+)\;B(K^+\p^-)}}{{\bf
p}_{K\p}}\;I(m_\p,m_K,\L)
\right\} \label{dm2_full}
\ey
where ${\bf p}_{ij}$ is the magnitude of the three-momentum in the
actual
decay and
\bz
I(m_1,m_2,\L)=-\int_{s_0}^{\L^2} \fr{\sqrt{1-\fr{s_0}{s}}\;ds}{
s-m_{D}^{2}} \label{int_def}
\ez
If the massless limit is taken in (\ref{dm2_full}) one recovers
(\ref{pp_ld})
with
the identification $\m^2=2\,m_D\,(\L-m_D)$. Although the result
depends
strongly on
the cutoff $\L$, this can be interpreted as the value of $s$ for
which
the internal momentum reaches its maximum. Not surprisingly, the
value of $\L$
giving an internal momentum of $\sim 1$~GeV is the same giving
$\m\sim 1$~GeV.
This is $\L\simeq (2-2.2)$~GeV, not too far above $m_D$.
Using this cutoff results in a contribution to the mass difference of
\bz
\fr{\D m_{D}^{\rm l.d.}}{\G}\simeq 6.5\times 10^{-4}\times
(1-\sqrt{b}))
\label{cha_est}
\ez
 As mentioned earlier, this result does not differ drastically
from what is obtained by using (\ref{pp_ld}).

On the other hand, it is clear that the use of (\ref{pp_ld}) to
estimate the
Pseudoscalar-Vector (PV) and Vector-Vector
(VV) contributions is dangerous and it could result in an
overestimate of these
contributions \cite{kaeding}.
 We can repeat the same procedure carried out  with the PP modes.
However data is even scarcer in these cases. Measurements for all the
PV and VV
decays rates as well as a more precise determination of $\G(D^0\to
K^+\p^-)$ is
needed in order to complete the picture of $\D m_D$ in this approach.
However,
and as it was already pointed out in \cite{dght86},
it is very likely that the
charged pseudoscalar contribution gives a good approximation to the
order of
magnitude of the effect. This is because,
although there are many contributions, their relative signs are not
fixed and
some degree of cancellation is expected. A hint of these
cancellations is
already present in (\ref{dm2_full}), which includes
 the interaction between the scale of softening of the effective
vertex and the
masses. It can be shown that in (\ref{dm2_full})
the contributions of small and large internal masses will tend to
have
have different signs. This is merely an argument that makes
cancellations
plausible, but by no means
a rigorous proof. It could be argued that several contributions might
conspire
to
give a total long distance contribution to
$\D m_D$ orders of magnitude larger than
(\ref{cha_est}). This scenario cannot be completely excluded in this
approach
until more data on PV and VV modes  is available.

\begin{center}
{\large Heavy Quark Effective Theory}
\end{center}

Yet another possible theoretical approach to $D^0$-${\bar D}^0$
mixing
is the application of the heavy quark effective theory (HQET). It was
first
noted in \cite{georgi} that if one considers the charm quark mass to
be
much larger than the typical scale of the strong interactions, there
would be no nonleptonic transitions to leading order in the resulting
effective theory. They would require large momenta to be exchanged
between
the heavy quark and the light degrees of freedom, a subleading effect
in inverse powers of the charm mass.
 As a consequence there are
no new available operators in the low energy theory to produce $\D
C=2$
transitions. At scales below $m_c$ these occur only due to operators
present
at the matching scale $m_c$ plus the action of the renormalization
group,
which in this picture constitute the only ``long distance'' effects.
Therefore,
to leading order in HQET, $\D m_D$ can  be computed from quark
operators. The
nonperturbative physics enters in the matrix elements of these
operators, and
in \cite{georgi,hqet2} are estimated using naive dimensional
analysis.
There are three groups of operators in the HQET.
The first corresponds to four-quark operators, which are the HQET
version of
the
box diagram. The second group, the six-quark operators, gives a
modest
enhancement
over the first one \cite{georgi,hqet2}. Finally, the eight-quark
operators give
a
large enhancement in the matrix elements over the four-quark
operators (a
factor
of $\approx 20$), but
they are suppressed by an overall factor of $\a_s/4\p$. This seems to
suggest
that the large enhancement in $\D m_D$ over the short distance box
diagrams
coming from individual contributions and accounted for in the
dispersive
approach above, is cancelled when all the contributions are summed
over in
order to make up for the $\a_s/4\p$ suppression \cite{georgi}.
 This is the HQET version of the
cancellations among different $SU(3)$ sets (e.g. PP with VV, etc.).
The size of the effect  is estimated in \cite{hqet2}, where QCD
corrections
and running are properly accounted for. The result is
\bz
\frac{\D m_D}{\G}\approx (1-2)\times 10^{-5} \label{hqe_est}
\ez
where the uncertainty comes from the unknown relative signs of the
various
operators. Thus HQET predicts a value of $\D m_D$ in the SM that is
roughly an
order of magnitude smaller than the dispersive estimates
(\ref{ld_esp}) and
(\ref{cha_est}).

Of course the validity of the central HQET assumption, $m_c\gg
\L_{QCD}$, can
be questioned. After all here $\L_{QCD}$ is actually a typical
hadronic scale,
not far below $1$~GeV. It is not clear what is the size of the
corrections.
However, the most interesting conclusion is the suggestion that there
is a
cancellation among the sets of $SU(3)$-related intermediate states.
On the other hand, the HQET and the dispersive results are consistent
within the experimental errors in the determination of $b$ in DCSD
modes.

\begin{center}
\large\bf
NEW PHYSICS AND $D^0-\bar{D}^0$ MIXING
\end{center}
Proposed high statistics charm experiments are likely to probe
$D^0-\bar{D}^0
$ mixing down to $r_D\sim 10^{-5}$ \cite{ch2000}.
 If the long distance contributions in the SM are below this
sensitivity, then
the question is: are there extensions of the SM that can fill this
window and
be compatible with all other low energy phenomenology ? In this
section we
review a few examples of new physics scenarios that could produce a
signal in
these experiments.

\begin{center}
\large\bf 1.~Two-Higgs Doublet Models
\end{center}

\begin{figure}
\vspace{8cm}
\includegraphics{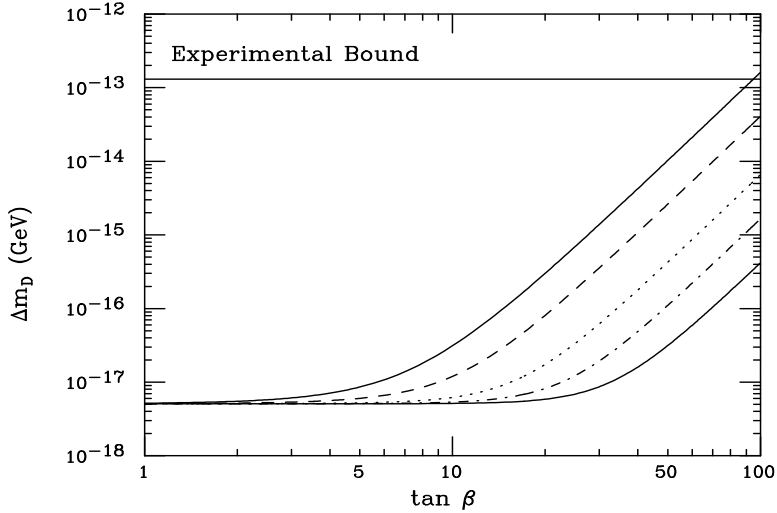}
\vspace{-1cm}
\caption[]{ $\D m_D$ in the two-Higgs doublet model~II as a function
of
$\tan\b$ and for several charged Higgs masses. From top to bottom
$m_{H^\pm}=50,100,250,500,1000$~GeV.}
\label{f_thd}
\end{figure}

As a first example of an extension of the SM we take a Two-Higgs
doublet model
with natural flavor conservation. That is, there are no tree level
FCNC
 \cite{thdm}.
 There
will be two neutral scalars, a neutral pseudoscalar and a pair of
charged
scalars. In what is called Model~II in the literature, the latter
couples to
the fermions as
\bz
{\cal L}=\fr{g}{\sqrt{2}m_W}\, H^+\left\{\cot\b \;\bar{\cal U}_R\;
M_u\; V_{\rm
ckm}\;  {\cal D}_L
 +\tan\b\;\bar{\cal U}_L\;V_{\rm ckm}\;M_d \;{\cal D}_R +{\rm
h.c.}\right\}
\label{thd_cou}
\ez
where ${\cal U}\equiv (u,\;  c,\; t)$, ${\cal D}\equiv (d,\;s,\;b)$,
$M_u$ and
$M_d$ are the
 diagonal quark mass matrices and $\tan\b\equiv v_2/v_1$; with
$v_1,\;v_2$ the
vacuum expectation values of the doublets. Incidentally, this type of
couplings
to fermions is the same as in the Higgs sector of the Minimal
Supersymmetric
Standard Model (MSSM), which is addressed separately below.
The couplings in (\ref{thd_cou})
induce an additional set of box diagrams where the $W^\pm$ is
replaced by the
charged Higgs. For large values of $\tan\b$ the $b$ quark is the
dominant
contribution, giving
\bz
\D m_{D}^{\rm 2HDM}\simeq
\fr{G_{F}^{2}}{6\p^2}\;m_D\;B_D\;f_{D}^{2}\;\h_{\rm
QCD}
\;|V_{cb}V_{ub}^{*}|^2\;m_{b}^{2}\;\tan^2\b\;F\left(\fr{m_{b}^{2}}{m_{
H}^{2}}\right) \label{thd_con}
\ez
where $\h_{\rm QCD}$ is a $QCD$ correction and $F(x)$ is a known
function of
the mass ratios, resulting from the loop integrals. As it is obvious
from
(\ref{thd_con}), the effect can be important for large $\tan\b$. This
shows
once again how charm meson physics can be complementary with $B$
physics. The
radiative decay $b\to s \g$ largely constrains the low $\tan\b$
region given
that, as it can be seen in (\ref{thd_cou}), the top quark mass term
amplifies
that region of parameter space. As seen in Fig.~\ref{f_thd}, in the
large
$\tan\b$ limit, charged Higgs masses below $m_H=250$~GeV will give a
contribution to which future experiments will be sensitive.

\begin{figure}
\vspace{8cm}
\includegraphics{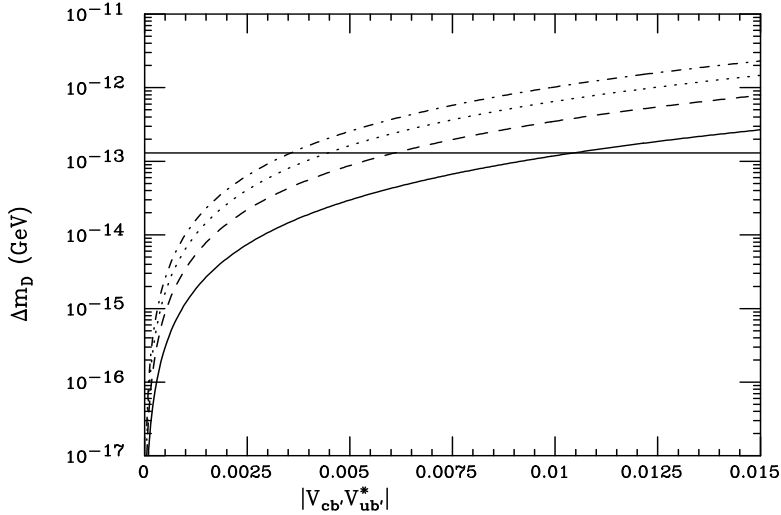}
\vspace{-1cm}
\caption[]{ The contributions to $\D m_D$ of a heavy $Q=-1/3$ quark,
as
a function of the mixing factor in a fourth-generation model. From
top to
bottom $m_{b'}=400,300,200,100$~GeV. The horizontal line shows the
current
experimental limit.}
\label{f_bpr}
\end{figure}

\begin{center}
\large\bf 2.~Heavy Down Quark
\end{center}
The reason why the SM box diagram contributions are
 so small is a very efficient GIM mechanism: the heaviest quark has
very small
couplings with $c$ and $u$ and is not so heavy anyway. The obvious
question is
then: what would happen if there was a fourth down quark in the loop
with
$Q=-1/3$ and a large mass. This could belong to a fourth generation
or be just
a singlet. Its contribution to the
mass difference is
\bz
\D m_{D}^{b'}\simeq
\fr{G_{F}^{2}m_{W}^{2}}{6\p^2}\;m_D\;B_D\;f_{D}^{2}\;\h_{\rm QCD}
\;|V_{cb'}V_{ub'}^{*}|^2\;F\left(\fr{m_{b'}^{2}}{m_{W}^{2}}\right)
\label{dmd_bpr}
\ez
{}From direct searches it is known that $m_{b'}>85$~GeV \cite{pdg94}.
 In the case of $b'$ belonging to a fourth generation, the couplings
to the
second and first generations are constrained by the possible
``leakage" of the
CKM matrix
from unitarity. The mixing factors must satisfy
\bz
|V_{ub'}|<0.08\quad ,\quad |V_{cb'}|<0.6 \label{bpr_upl}
\ez
although smaller mixing factors are expected. The effect of the $b'$
quark is
shown
in Fig.~\ref{f_bpr}. If the mixing factors are not too small a heavy
$b'$
could give a large effect, saturating the current experimental limit.

\newpage
\begin{center}
\large\bf 3.~Tree Level FCNC
\end{center}
Tree-level FCNC are severely constrained by $K^0-\bar{K}^0$ and
$B^0-\bar{B}^0$
mixings and decays like $K^+\to\p^+\n\bar\n$ and $B\to X\ell^+\ell^-$
\cite{pdg94}.
However, it is possible to imagine scenarios where the up-quark
sector
is treated differently. Some of these new physics scenarios predict
large
$D^0-\bar{D}^0$ mixing effects.

A first example is to relax the requirement of flavor conservation in
multi-Higgs models. There will be tree-level flavor changing
couplings of the
neutral Higgses \cite{pak_sug}.
 The form of the flavor conserving couplings to fermions
suggests the parametrization of the flavor changing couplings given
by
\bz
C_{ij}\sim \fr{\sqrt{m_i\;m_j}}{v_{\rm w}}\; \d_{ij} \label{fch_cou}
\ez
where $i,j$ are the quark labels (e.g. $u, c, t$); and $v_{\rm
w}\simeq
246$~GeV. The $c-u$ coupling would then induce a contribution to the
mass
difference of the form
\bz
\D m_{D}^{h^0}=\fr{5}{12}\;f_{D}^{2}\;B_D\;m_D\;\fr{m_cm_u}{v_{\rm
w}^{2}}
\fr{\d_{cu}^{2}}{m_{h}^{2}} \label{fch_con}
\ez
where $m_u$ is the current quark mass. The experimental limit is
saturated
for reasonable values of $\d_{cu}\sim (0.2-1)$ and $m_h\sim
(50-250)$~GeV.

Another interesting example is that of theories of dynamical symmetry
breaking with large GIM violations, such as TopColor models
\cite{topcolor}.
 Here the top quark
gets a large mass after the breaking down to $QCD$ of a group
strongly coupled
to the third generation. This generates a set of pseudo-Goldstone
bosons
(top-pions) that couple to the up-quark sector. After the quark weak
eigenstates are rotated to the mass eigenstates, there will be flavor
changing
neutral
couplings mediated by the top-pions of the form
\bz
C_{ij}\sim \fr{m_t}{\sqrt{2}\;f_{\tilde{\p}}}\;U_{L}^{ti}\;U_{R}^{tj}
\label{top_col}
\ez
where $f_{\tilde{\p}}\simeq 50$~GeV is the top-pion decay constant
and $U_L$
and $U_R$ are the mass matrices of the
 left and right up-quark sector respectively. These couplings lead to
a mass
difference given by
\bz
\D m_{D}^{\rm
t.c.}\simeq\fr{5}{12}\;f_{D}^{2}\;B_D\;m_D\;\fr{m_{t}^{2}}{2f_{\tilde{
\p}}^{2}}\;
\fr{U_{L}^{tc}U_{R}^{tu}U_{L}^{tu}U_{R}^{tc}}{m_{\tilde\pi}^{2}}
\label{toc_con}
\ez
The mass matrices are not determined in general by the model but must
obey, as
in the SM, $V_{\rm ckm}=U_{L}^{\dagger}\;D_R$. This suggests that a
possible
anzatz for $U_{L,R}$
is to take the ``squared root"
 of $V_{\rm ckm}$. This prescription gives, for instance,
$U_{L}^{tc}\simeq
(1/2)\;V_{cb}$. Following this in (\ref{toc_con}) and taking
$m_{\tilde
\p}\simeq 200$~GeV \cite{topcolor},
gives $\D m_{D}^{\rm t.c.}\simeq 8\times 10^{-14}$~GeV, right below
the current
experimental limit. In principle, other textures can be chosen that
would
not give such a large effect.

\begin{center}
\large\bf 4.~Supersymmetry
\end{center}
In addition to the box diagrams involving charged Higgses,
there will be contributions from squarks~$+$~(gluinos, charginos or
neutralinos) box diagrams. These  vanish if the down squarks are
degenerate.
This is just
a statement derived from GIM cancellations. However, there could be
flavor
changing, radiatively generated mass insertions \cite{hagelin}.
 They are thought to be small
in the MSSM, but could be large in non minimal models \cite{nirsusy}.
 For instance, the action of these mass insertions would allow for
squark-gluino box diagrams. The resulting $\D C=2$ hamiltonian is
\by
{\cal H}_{\D C=2}^{\rm SUSY}&=&\fr{\a_{s}^{2}}{216\tilde{m}_0}\left\{
\fr{\d\tilde{m}_{u_Lc_L}^{2}}{\tilde{m}_{0}^{2}}\;
G\left(\fr{m_{\tilde{g}^{2}}}{\tilde{m}_{0}^{2}}\right)\left(\bar{u}_L
\g_\m
c_L\right)\left(\bar{u}_L\g^\m c_L\right) \right. \nn \\
& &\left.\quad\quad +(RR)^2+(LL)(RR)+(LR)^2+(LR(LR)\right\}
\label{sus_con}
\ey
where $\tilde{m}_0$ is the universal scalar mass, $m_{\tilde{g}}$ is
the gluino
mass,
$\d\tilde{m}_{u_Lc_L}$ characterizes the mass insertion and the
$G(x)$'s  are
known functions. This way, the experimental upper limit on $\D m_D$
can be
translated into limits for the various terms entering in
(\ref{sus_con}).
These are
\bz
\fr{\d\tilde{m}_{u_Ac_B}^{2}}{\tilde{m}_{0}^{2}} < \left\{
\ba{cl}
(0.2)^2 & {\rm ,~for~} (LL)^2, (RR)^2 \\
(3.6\times 10^{-2})^2 & {\rm ,~for~} (LL)(RR) \\
(5\times 10^{-2})^2&{\rm ,~for~} (LR)^2 , (RL)^2 \\
(0.1)^2& {\rm ,~for~} (LR)(LR)
\ea
\right.
\ez
Thus, the current experimental limit is already sensitive to non
minimal SUSY
effects.

There are several other extensions of the SM that would saturate the
current
experimental limit or at least give a signal in high statistics charm
experiments. The above list is by no means exhaustive and is mainly
intended
to illustrate how the effect comes about in a variety of theories.

\vskip 1cm
\begin{center}
\large\bf
CP VIOLATION
\end{center}
The $D$ system is not particularly sensitive to CP violation in the
SM to the
extent the $K$ and $B$ mesons are. Once again, this could imply there
is a
window of observation of new physics effects. In what follows we
discuss some
of the general features of CP violation in $D$ mesons rather than
going into
specific calculations, both in the SM and beyond.

\begin{center}
\large\bf Direct CP Violation
\end{center}
The occurrence of direct CP violation requires the concurrence of
both weak and
strong relative phases between two or more amplitudes contributing to
a given
final state. In the SM, relative weak phases can only be obtained in
Cabibbo
suppressed decays, for instance,  via the interference between
spectator and
penguin amplitudes. To estimate the size of the CP asymmetries this
would
generate, we write
\by
a_{\rm CP}&\sim&
\fr{Im[V_{cd}\;V_{ud}^{*}\;V_{cs}\;V_{us}^{*}]}{\l^2}\;\;
\sin\d_{\rm st}\;\;\fr{P}{S} \label{asy_dir} \\
&\sim &A^2\;\h\;\l^4\;\sin\d_{\rm st}\;\fr{P}{S} \quad\quad\leq
10^{-3} \nn
\ey
where $\d_{\rm st}$ is the strong relative phase between the penguin
and the
spectator amplitudes, and $A\sim 1$ and $\h$ are CKM parameters in
Wolfenstein's parametrization.
Specific model calculations for $D\to KK, \p\p, K^*K$, three-body
modes, etc.
yield this order of magnitude for the effect. New physics could
enter, for
instance, through large phases in the penguin diagram. This could
give
very large asymmetries of the order of one percent or larger.
On the other hand, an even cleaner window are the Cabibbo allowed
decays. These
modes do not have two amplitudes with different weak phases and
therefore the
CP asymmetry is zero in the SM. There are new physics scenarios that
provide
extra phases and could give asymmetries as large as one percent. This
is for
instance the case in some left-right symmetric models \cite{left_r}.
 The current experimental sensitivities for various modes is in the
vicinity
if $10\%$ \cite{cpexp}.

\newpage
\begin{center}
\large\bf Indirect CP Violation
\end{center}
The interaction between mixing and  CP violation in $D$ mesons
has recently received a lot of attention in the literature
\cite{blaylock,wolf95,brow_pak}.
 Here we shall
focus only on one aspect, which can be condensed in the following
question:
if mixing is large (e.g. right below the current experimental limit)
should CP
violation in $D$ decays be large ?
The question is motivated by the fact that in $B$ decays the large
$B^0-\bar{B}^0$ mixing is known to give large CP asymmetries.
We first define the time-evolved states in the usual way, as
\by
|D^0(t)\rangle &=&g_+(t)\;|D^0\rangle +
\fr{q}{p}\;g_-(t)\;|\bar{D}^0\rangle
\nn \\
|\bar{D}^0(t)\rangle &=&\fr{p}{q}\;g_-(t)\;|D^0\rangle +
g_+(t)\;|\bar{D}^0\rangle \nn
\ey
with
\bz
\fr{p}{q}\equiv \sqrt{\fr{M_{12}-i\G_{12}}{M_{12}^{*}-i\G_{12}^{*}}}
\label{poq_def}
\ez
and the time evolution given by
\bz
g_{\pm}(t)=\fr{1}{2} e^{(-\fr{\G_L}{2}t+im_Lt)}\left[1\pm
e^{(-\fr{\D\G_D
t}{2}+i\D m_D t)}\right] \label{tim_evol}
\ez
We also need to define  the amplitudes
\by
A&\equiv& \langle f|H_{w}| D^0\rangle \quad\quad ;\quad\quad
B\equiv\langle
f |H_w |\bar{D}^0\rangle  \nn \\
\bar{A}&\equiv& \langle \bar{f}|H_{w}| \bar{D}^0\rangle \quad\quad
;\quad\quad
\bar{B}\equiv\langle \bar{f} |H_w| D^0\rangle  \nn
\ey
and the ratio
\bz
\bar{\rho}\equiv \fr{A}{B}\label{rho_def}
\ez
To simplify notation we consider the case when $f=CP$~eigenstate
(e.g. $\p\p$,~
$KK$, etc).
If mixing is a large effect, let us say right below the current upper
limit,
and therefore due to
new physics contributions then it is very likely that
 $\D m_D/\G \gg \D\G_D/\G$ is a very good approximation: non standard
contributions to $\D\G_D$ are constrained by actual branching ratios.
Under
these assumptions the asymmetry takes the form
\bz
a_{\rm CP}\simeq \fr{1-\left|\fr{q}{p}\bar{\r}\right|^2-2\fr{\D
m_D}{\G}
Im\left[\fr{q}{p}\bar\r\right]}{\left(1+\left|\fr{q}{p}\bar{\r}\right|
^2\right)\;
\left(1+(\fr{\D m_D}{\G})^2\right)} \label{acp_mix}
\ez
Small direct CP violation implies $|(q/p)\bar{\r}|^2\simeq 1$. This
leads to
\bz
a_{\rm CP}\simeq \fr{\D m_D}{\G}\times Im\left[\fr{q}{p}\bar\r\right]
\simeq\fr{\D m_D}{\G}\; 2\h A  \label{acp_sm}
\ez
where the last step follows in the SM and from considering the

contribution of the $b$ quark to the imaginary part of the box
diagram.

The resulting asymmetry can be then of the order of $\D m_D/\G$ even
if
only the SM phases intervene. However in models giving large mass

differences it is also likely that there will be additional

CP violating phases.

 These  could be present
in the {\em non} SM contributions to the mass difference and they
mostly affect
 $q/p$ and not $\bar\r$. Several non standard scenarios for
generating
these additional phases are discussed in \cite{blaylock}.

\vskip 1cm
\begin{center}
\large\bf
RARE AND RADIATIVE DECAYS
\end{center}
Let us first address the distinction made between rare and radiative
decays.
Radiative weak decays of charm mesons do no effectively test the SM.
To see
this let us take the transitions governed by the short distance
flavor
changing vertex $c\to u\g$. In the SM they occur only at one loop
through the
electromagnetic penguin, analogous to $b\to s\g$. However in this
case, the
inclusive branching ratio, even after very large $QCD$ corrections,
is very
small:
$B(c\to u\g)\simeq 10^{-12}$ \cite{bghp1}.
This, however, does not constitute a window for new physics. There
are more
mundane contributions to the corresponding exclusive processes, like
$D^0\to
\r^0\g$, that do not arise from short distance physics. These ``long"
distance
contributions can be thought of as
coming from either pole diagrams (arising from quark exchange) and
vector meson
dominance diagrams, all of  which are not calculable from first
principles but
can be
estimated in models to give $B(D^0\to\r^0\g)\simeq (1-5)\times
10^{-6}$
\cite{bghp1}.
 This large rates preclude the use of these modes as SM tests. On the
other
hand, a
better theoretical understanding of these modes is interesting in its
own right
as well as in order to understand possible
long distance contamination in radiative $B$ decays \cite{ld_inb}.
The availability of several decay modes -$D^0\to \r^0\g$, $D^0\to
\bar{K}^*\g$,
$D_s\to\r^+\g$, etc.- at branching fractions of ${\cal O}(10^{-6})$
or larger
will improve our knowledge of strong dynamics at the charm scale
\cite{bigi}.

Truly rare decays are those whose SM rates are extremely small or
simply zero.
Most of them proceed through FCNC induced at one loop in the SM. The
simplest
example is $D^0\to\ell^+\ell^-$, with $\ell=e$ or $\m$. Their
branching ratios
are smaller than $10^{-15}$, even after long distance contributions
are
taken into account. There are experimentally clean and any signal in
any of
these channels would imply new physics. However the helicity
suppression is a
factor of $\sim 10^{-3}$ in the case of $\m$, and $10^{-7}$ for the
$e$. Modes
without this suppression, like $D\to X\ell^+\ell^-$ and  $D\to
X\nu\bar\nu$,
are more likely to show the first signals if, for instance,  there is
a new
physics mechanism underlying the short distance transition $c\to u$.
In the SM their branching ratios are expected to be of ${\cal
O}(10^{-8})$ or
smaller in the charged lepton cases and negligibly small in the
neutrino modes.
New
upper limits in the $D^+\to\p^+\ell^+\ell^-$ channels have been
recently
reported  by E791 \cite{e791}:
$B_{\m\m}<1.8\times 10^{-5}$ and $B_{ee}<6.6\times 10^{-5}$.
These are already constraining extensions of the SM in a way
complementary with
mixing. Predictions for new physics scenarios can be found in several
places in
the literature \cite{lepqua}, but it is necessary to update them and
most
importantly to
take into account other pieces of phenomenology now available.

\vskip 1cm
\begin{center}
\large\bf
CONCLUSIONS
\end{center}
We have seen that charm meson physics offers several opportunities
to observe  the effects of new physics. This is mainly due to the
suppression of the signals in the SM.
In the case of $D^0-\bar{D}^0$ mixing, the estimate of the long
distance
SM contribution is very uncertain. However, the question of interest
at the
moment is whether
\bz
r_{D}^{\rm SM}<10^{-5} \label{sm_uplim}
\ez
is a correct upper limit. This is relevant because is the planned
sensitivity
of future high statistics charm experiments \cite{ch2000}.
 If $r_{D}^{SM}$ is $10^{-10}$ or
$10^{-7}$ is an issue theorists should worry about, but it will not
affect the
interpretation of the outcome of these experiments. The validity of
(\ref{sm_uplim}) is not a settled question among theorists. We have
seen two
approaches to $r_{D}^{\rm SM}$ satisfying (\ref{sm_uplim}): a
dispersive
approach and the HQET approach. However, these are approximate
calculations
and there are those who point out that  (\ref{nai_est}) with a unit
coefficient is another way of estimating the effect \cite{wolf86}.
The high end of this
estimate violates (\ref{sm_uplim}). The theoretical community should
make an
effort to resolve this outstanding problem. More data in nonleptonic
$D$ decays
is needed in order to see if there is a pattern of cancellations as
suggested
by the HQET. In the meantime, experiments might provide with an
independent way
of deciding on the origin of an observation of $r_D$: the direct
measurement
of the lifetime difference $\D\G_D$. This can be done, assuming CP
conservation, by  looking at the decay time distribution of $D$
decays to
CP even and odd final states \cite{liu}.
The difference of the slopes is proportional to
$\D \G_D$. This quantity is not prone to get contributions from new
physics but
rather to be entirely given by the SM: after all it is a sum over
real
intermediate states. Moreover, it should be of the same order of
magnitude as
the long distance contributions to $\D m_D$. Thus, not only would
this allow
the separation of $\D m_D$ and $\D\G_D$ but also, for instance, point
at new
physics if $\D m_D$ is observed and $\D\G_D$ is not seen at the
corresponding
level. Considering how many extensions of the SM saturate the
experimental
limit, this program makes $D^0-\bar{D}^0$ mixing an important window
for new
physics in the future.

With respect to CP violation,  we have seen that  asymmetries at the
level of
one percent would signal new physics. We also pointed out that CP
violation due
to mixing will be enhanced by new physics only if both $\D m_D$ and

the entering phases, both from the SM and/or new physics, are large.

For rare decays a lot more work is needed in order to establish what
level of
branching ratios are allowed in each new physics scenario once all
the
constraints from low energy phenomenology are factored in.

\vskip 1cm
\begin{center}
\large\bf
AKNOWLEDGEMENTS
\end{center}
This presentation reflects work in collaboration with Eugene
Golowich, JoAnne
Hewett and Sandip Pakvasa. I would also like to thank Jose Repond and
the rest
of the organizers of the workshop for their excellent effort. This
work was
supported by the U.S. Department of Energy.

\end{document}